\documentstyle[multicol,epsf,aps,prl,twoside]{revtex} 
\begin{document}

\title{Evidence for softening of first-order transition in $3D$ by 
        quenched disorder}

\author{Christophe Chatelain$^{1,2}$, Bertrand Berche$^1$, 
        Wolfhard Janke$^2$, and Pierre Emmanuel Berche$^3$}
\address{$^1$Laboratoire de Physique des Mat\'eriaux,~\cite{byline2} 
        Universit\'e Henri Poincar\'e, 
        Nancy 1,\\ BP 239,
        F-54506  Vand\oe uvre les Nancy Cedex, France\\ 
        $^2$Institut f\"ur Theoretische Physik, Universit\"at Leipzig, 
        D-04109 Leipzig, Germany\\
        $^3$Groupe de Physique des Mat\'eriaux,~\cite{byline3}
        Universit\'e de Rouen, F-76821 Mont Saint-Aignan Cedex, France}
       
\date{\today}
\maketitle

\begin{abstract}
We study by extensive Monte Carlo simulations the effect of random bond dilution
on the phase transition of the three-dimensional 4-state Potts model which is
known to exhibit a strong first-order transition in the pure case. The phase 
diagram in the dilution-temperature plane is determined from the peaks of the 
susceptibility for sufficiently large system sizes. In the strongly disordered 
regime, numerical evidence for softening to a second-order
transition induced by randomness is given. Here a large-scale finite-size scaling
analysis, made difficult due to strong crossover effects presumably 
caused by the percolation fixed point, is performed. 
\end{abstract} 

\vspace*{-9mm}
\begin{multicols}{2}
\narrowtext

The influence of random, confining geometries on first-order phase transitions
has been the subject of exciting experimental studies in the past few years.
The case of the isotropic to nematic transition of $nCB$ liquid crystals
confined into the pores of aerogels consisting of multiply connected
internal cavities has been particularly extensively studied and led to 
spectacular results: The first-order transition of the corresponding bulk
liquid crystal is drastically softened in the porous glass and becomes
continuous~\cite{IannacchioneEtAl93}, an effect which was not attributed to 
finite-size effects but rather to the influence of random disorder.

The first attempt to reproduce such a softening scenario using Monte Carlo (MC)
simulations was reported by Uzelac {\em et al.\/}~\cite{UzelacHasmyJullien95}
who studied a three-dimensional (3D) $q$-state Potts model with spin variables
(taking $q=3$ and $4$ states per spin) located inside the randomly connected 
pores of an aerogel modeled by diffusion-limited cluster aggregation. Although
in experimental studies~\cite{WuEtAl92etc} commonly random fields or random 
uniaxial anisotropies are suggested to explain the softening of the transition,
the random disorder chosen in Ref.~\onlinecite{UzelacHasmyJullien95} is coupled
to the energy density and thus more akin to bond-dilution.

The qualitative effect of random bond disorder on
second-order phase transitions is well understood through the Harris relevance 
criterion \cite{Harris74}, and a beautiful experimental confirmation was 
reported in a LEED investigation of a two-dimensional (2D) order-disorder 
transition~\cite{SchwengerEtAl94}. For systems with a first-order transition
in the pure case, randomness generically softens the transition and, under 
certain circumstances, may even induce a second-order transition according 
to a picture first proposed by Imry and Wortis~\cite{ImryWortis79etc}.

In 2D, the natural candidate for theoretical investigations is the
$q$-state Potts model, since this model is exactly known to exhibit regimes
with first- and second-order transitions~\cite{Wu82}, depending on the value
of $q$. Many results were obtained in both regimes in the last ten 
years~\cite{Ludwig87etc,Derrida84etc}, including approximate analytic 
treatments, MC simulations, transfer-matrix calculations, and high-temperature
series expansions. Among others also quite intricate problems such as 
self-averaging and multi-fractality have 
recently been studied in some detail \cite{Derrida84etc}. 
 
In 3D, to date only the Ising model with site dilution 
has been studied extensively~\cite{WangChowdhury89etc}. 
In accordance with the Harris criterion the presence of random disorder was 
found to modify the critical exponents to values close to $\nu=0.684(5)$,
$\gamma/\nu=1.96(3)$, and $\beta/\nu=0.519(8)$. Concerning the influence of 
random disorder on first-order transitions, even less is known in 3D. Apart 
from the exploratory work~\cite{comment} of 
Uzelac {\em et al.\/}~\cite{UzelacHasmyJullien95},
only the site 
diluted 3-state Potts model, which in the pure 3D case has a very weak 
first-order transition \cite{JaVi97}, has recently been studied via 
large-scale MC simulations~\cite{BallesterosEtAl00}. This study led to the 
conclusion that the critical exponent $\nu$ governing the scaling behavior 
of the correlation length is compatible with that of the 3D site diluted 
Ising model, whereas the $\eta$ exponent is definitely different.
 
The purpose of this Letter is to present numerical evidence for softening of 
the transition when it is {\em strongly\/} of first order in the pure system,
in order to be sensitive to disorder effects. The paradigm in 3D is the 4-state
Potts model, since the correlation length at the transition temperature of the
unperturbed system is small enough ($\xi\simeq 3$ in lattice spacing
units~\cite{JankeKappler96}) to allow simulations of significantly large 
systems. For the pure 3D 5-state Potts model the first-order transition is 
already too strong~\cite{JankeKappler96}.
In the following we, therefore, consider the
4-state bond diluted Potts model on simple-cubic lattices of size $V = L^3$ 
with periodic boundary conditions. 
The Hamiltonian of the system with independent, quenched random interactions is 
written as $-\beta {\cal H}=\sum_{(i,j)}K_{ij}\delta_{\sigma_i,\sigma_j}$, 
where the spins take the values $\sigma_i=1,\dots,4$ and the sum goes over 
all nearest-neighbor pairs $(i,j)$. The coupling strengths are allowed to take 
two different values $K_{ij}= K \equiv J/k_BT$ and $0$ with probabilities $p$ 
and $1-p$, respectively. The order parameter for a given realization of the 
$K_{ij}$ is 
defined by the majority orientation of the 
spins,
$m = \langle\mu\rangle$, where $\mu={(q\rho_{\rm max}-1)}/{(q-1)}$
and $\rho_{\rm max}$ is the maximum value of the density of spins $\rho_\sigma$
in the $q = 4$ possible spin states. The thermal average over the MC iterations
is indicated by brackets $\langle \dots \rangle$, and the physical quantities 
are then averaged over disorder realizations, e.g.,  
$\bar m = \overline{\langle\mu\rangle}$. For each disorder realization the 
susceptibility is defined as usual via the fluctuation-dissipation theorem, 
$\chi = K V (\langle\mu^2\rangle-\langle\mu\rangle^2)$.


The present MC simulations consist of two parts. First, a scan of the 
dilution-temperature plane in order to determine the phase diagram, and second,
a large-scale finite-size scaling (FSS) study at $p=0.56$ up to $L=96$ in the 
dilution regime exhibiting second-order transitions. The spin updates were 
performed with the cluster-flipping method~\cite{SwendsenWang87} 
in the Swendsen-Wang  formulation
which turned out to be better behaved than the Wolff single-cluster 
version for high dilutions (small $p$), where small 
clusters connected by non-vanishing bonds
are more likely to appear. We made sure that, by adjusting the 
length of the runs of cluster algorithms, at least 10 tunnelling 
events between the two 
coexisting phases of 
the pure system ($p=1$) were observed up to lattice size $L=16$.
To improve the accuracy, 
for weak dilutions ($p$ between 1 and 0.68) 
where we obtained evidence for first-order transitions,
we also used  multicanonical 
algorithms~\cite{Janke94}.  
For the determination of the maxima of observables we applied standard 
histogram reweighting 
techniques 
in order to 
extrapolate the results over a temperature range around the simulation point.

At each probability $p$ and for each realization of the random couplings, 
between $15 \times 10^3$ to $30 \times 10^3$ MC sweeps per spin were performed,
resulting in at least 250 (almost) independent measurements of the physical 
quantities for 
the largest lattice size considered. This turned out to be sufficient for 
reliable thermal averages. 
For the average over disorder 
realizations, between 2\,000 and 5\,000 samples were generated.

\begin{figure}
        \epsfysize=6.5cm
        \begin{center}
        \mbox{\epsfbox{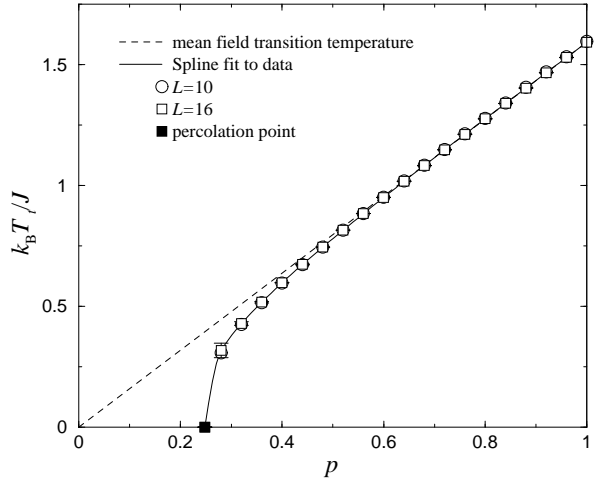}}
        \end{center}\vskip 0cm
        \caption{Phase diagram of the bond diluted 4-state Potts model in
        3D. The solid line is a spline interpolation to guide the eye, and 
        the dashed line shows the transition line within a simple
        mean-field argument, $T_t(p)=p T_t(1)$.}
        \label{FigPhaseDiag}  
\end{figure}

The phase diagram is determined from the locations of the maxima of the average
susceptibility, $K_{\rm max}$, obtained for systems of increasing sizes up 
to $L=16$. Theoretically we expect that the transition remains of 
first order in the regime of low impurity concentrations. For increasing 
concentrations a regime of second-order transitions should appear from a 
\mbox{(tri-)} critical concentration until the percolation threshold 
is reached where the transition vanishes altogether.
After this percolation threshold, no ordered phase can exist at any finite
temperature. Two points are known in the $p-k_BT/J$ plane: The
transition temperature $T_t$ of the pure system~\cite{JankeKappler96}
$       k_BT_t(p=1)/J\simeq 1.59076$,
and $   k_BT_t(p=p_c)/J=0$
at the percolation threshold $p_c \approx 0.2488$. As can be inspected in
Fig.~\ref{FigPhaseDiag}, the temperatures of the susceptibility maxima for
different lattice sizes are very stable and already for $10^3$ spins an accurate
transition line is obtained. Also shown is the result of a
simple mean-field argument taking into account the average number of
neighbors, $k_BT_t(p)/\bar zJ=\rm const$, where the constant is chosen such
that $T_t$ of the pure system is reproduced. This leads to a simple linear
approximation of the transition line, $k_BT_t(p)/J=1.59076\times p$, which is
surprisingly accurate over a significant range of $p$-values. 

By monitoring the FSS behavior of various thermodynamic quantities
as well as the
(pseudo-) dynamics of the update algorithm, we estimate the
tricritical point to be located between $p=0.68$ and $0.84$:
The shape of the energy probability densities as well as the Binder 
cumulants suggest a first-order transition at $p=0.84$ and above while a
clear second-order signal is observed at $p=0.68$ and below.
For the investigation of the critical properties in the second-order regime
we have chosen a dilution $p=0.56$ where the corrections to scaling are 
seemingly relatively small, since the effective transition temperatures
corresponding to the susceptibility maxima remain almost constant in
the range of sizes $2\leq L\leq 16$ used for the determination of the phase
diagram.

\begin{figure}[t]
        \epsfysize=6.4cm 
        \begin{center}
        \mbox{\epsfbox{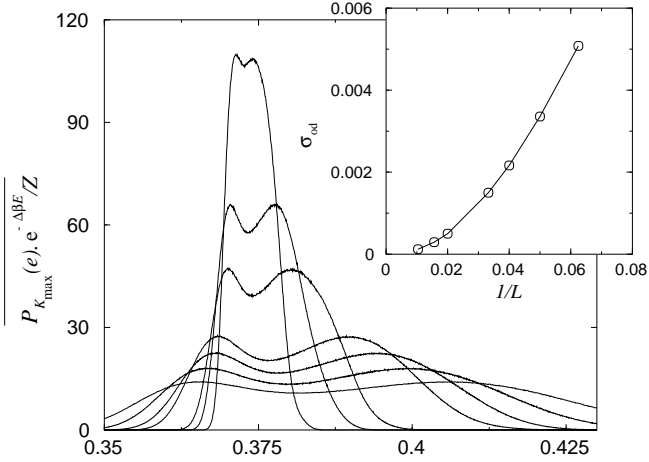}}
        \end{center}\vskip -0.2cm
        \caption{The probability density of the
        energy at $p=0.56$ for
        sizes $L=16$, 20, 25, 30, 50, 64, and 96, reweighted to
        the temperature where the two peaks are of equal height. The
        inset shows the associated effective interface tension.}
        \label{FigP_de_E_reweight}
\end{figure}

In order to convince ourselves that for $p=0.56$ the transition is indeed
of second order, let us first consider the average probability densities 
of the energy, $P(e)$. In Fig.~\ref{FigP_de_E_reweight} their shapes close to 
$K_{\rm max}$ are depicted for various lattice sizes up to $L=96$. We see that 
the system exhibits for small sizes two distinct peaks which clearly collapse 
into a single peak when one approaches the thermodynamic limit. 
This is precisely what is expected at a second-order phase transition, while in 
the case of a first-order transition the double-peak structure should persist
for all sizes and, in fact, should become even more pronounced when the system 
size increases. Physically, a two-peak structure would reveal the presence of two 
phases at the transition temperature, and this coexistence is the characteristic feature
of a first-order transition. 
More quantitatively we have confirmed that the effective interface tension 
$\sigma_{\rm od}$ derived from the usual relation 
$P_{\rm min}/P_{\rm max} \sim \exp(- 2\sigma_{\rm od}(L) L^{d-1})$
vanishes in the infinite-volume limit; see the inset of 
Fig.~\ref{FigP_de_E_reweight}.

\begin{figure}[t]
        \epsfysize=6.4cm
        \begin{center}
        \mbox{\epsfbox{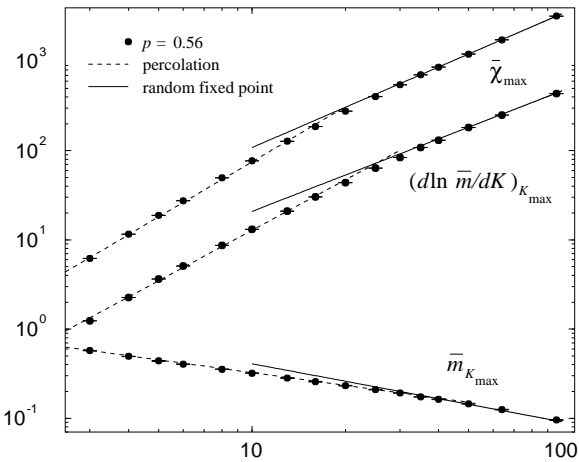}}
        \end{center}\vskip -0.3cm
        \caption{Finite-size scaling behavior of the susceptibility, the
        magnetization and of $d\ln\bar m/dK$ at $K_{\rm max}$ (the quantities
        have been shifted in the vertical direction for the sake of clarity).
        The scaling behavior for small lattice sizes is
        presumably governed by the percolation fixed point, and above a
        crossover length scale it reaches a new (random) fixed point.}
        \label{FigChiMaxM}
\end{figure}

Our final goal is a quantitative characterization of the critical behavior by
providing estimates for the critical exponents of the transition. To this end
we have performed a standard FSS analysis at $p=0.56$. As can be inspected in
Fig.~\ref{FigChiMaxM}, the corrections to asymptotic FSS
for $\bar\chi_{\rm max}$ seem to become quite small
above $L=30$. The data are thus linearly fitted to $a_\chi L^{\gamma/\nu}$ for
$L$ in the ranges $L_{\rm min} > 30$ to $L_{\rm max}=96$, and the resulting exponents are
collected in Table~\ref{TabApp1}.
Selecting the fits with the smallest chi-squared per degree of freedom,
$\chi^2/{\rm d.o.f}$, we take as the final result
the lines in bold face, e.g., $\gamma/\nu=1.50 \pm 0.02$.
%
%
\vspace*{-0.5cm}
\vbox{
        \narrowtext
        \begin{table}
        \caption{Linear fits for $\bar\chi_{\rm max}$,$\!$
        $(\partial_{K}\ln\bar m)_{K_{\rm max}}$,$\!$ and
        $\bar m_{K_{\rm max}}$.
        \label{TabApp1}}
        \begin{tabular}{llll} 
        $L_{\rm min}$ & $L_{\rm max}$ & $\gamma/\nu$
        & $\chi^2/{\rm d.o.f}$ \\
        \tableline 
\bf    35  &\bf  96  &\bf 1.500(14) &\bf 0.044 \\
       40  &     96  &    1.502(17) &    0.054 \\
       50  &     96  &    1.506(27) &    0.065 \\
        \tableline
        \tableline
        $L_{\rm min}$ & $L_{\rm max}$ & $1/\nu$
        & $\chi^2/{\rm d.o.f}$ \\
        \tableline
     35  &     96  &     1.362(13)  &    1.011  \\
     40  &     96  &     1.353(16)  &    0.887  \\
\bf  50  &\bf  96  &\bf  1.330(25)  &\bf 0.419  \\
        \tableline
        \tableline
        $L_{\rm min}$ & $L_{\rm max}$ & $\beta/\nu$
        & $\chi^2/{\rm d.o.f}$ \\
        \tableline
     35  &     96  &     0.592(13)  &    2.778  \\
     40  &     96  &     0.608(15)  &    2.145  \\
\bf  50  &\bf  96  &\bf  0.645(24)  &\bf 0.311  \\
        \end{tabular}
        \end{table}
        \narrowtext
}\vskip -0mm
\noindent

The quantity $\partial_K\ln \bar m$ gives an estimation of the exponent $\nu$,
$(\partial_{K}\ln \bar m)_{K_{\rm max}} \propto L^{1/\nu}$. Here our analysis
leads to an
estimate of $1/\nu=1.33\pm 0.03$ or $\nu=0.752\pm 0.014$, in agreement with the
stability condition
of the random fixed point ($\nu \ge 2/D =0.666\dots$)
and significantly different from the estimate for the site diluted
3D Ising model ($\nu = 0.684(5)$).
The same procedure was applied to the magnetization evaluated at the
temperature where the susceptibility is maximal.
Judging the values of $\chi^2/{\rm d.o.f}$ would lead to the result
given in bold face in Table~\ref{TabApp1}, but the
effective exponent for the magnetization is clearly not yet stable. 
We therefore also considered the FSS behavior of higher (thermal) moments of 
the magnetization, $\overline{\langle\mu^n\rangle}$ which should scale with a
dimension $n\beta/\nu$. The results for the first moments
exhibit, however, again much stronger corrections to scaling than we
observed for $\bar\chi$ or $\partial_K\ln \bar m$, leading to our final
estimate of $\beta/\nu=0.65 \pm 0.05$. . 

\begin{figure}[b]
        \epsfysize=5.6cm
        \begin{center}
        \mbox{\hspace{-0.01cm}\epsfbox{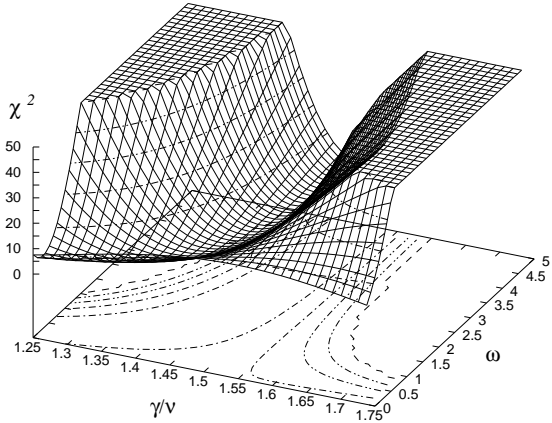}}
        \end{center}
        \caption{Plot of the $\chi^2$ deduced from linear fits of
        $\bar\chi_{\rm max}(L)=a_\chi L^{\gamma/\nu}(1+b_\chi L^{-\omega})$
        in the range $25\leq L\leq 96$.
        The exponents are fixed parameters and the amplitudes are
        free. The base plane gives the ranges of variation of the exponents:
        $1.25\leq\gamma/\nu\leq 1.75$ and $0\leq\omega\leq 5$.
        A cutoff at $\chi^2=50$ has been introduced
        for clarity of the figure.
        }
        \label{FigChi2}
\end{figure}

From the log-log plots of the three quantities in Fig.~\ref{FigChiMaxM}
one can clearly observe a crossover from a percolation-type behavior
at small sizes, characterized by the exponents~\cite{LorenzZiff97}
$\gamma/\nu\simeq 2.05$, $1/\nu\simeq 1.124$, and $\beta/\nu\simeq 0.45$,
towards a new regime at large sizes, which presumably
corresponds to the random fixed point, with exponents as given above.
The numerical evidence for this interpretation is quite striking, but with the
present system sizes we can of course not completely rule out the possibility
of corrections to scaling, in particular since for the 3D disordered Ising model
it is well known that such corrections at the random fixed point are strong (with a
correction-to-scaling exponent around $\omega=0.4$).
In order to investigate this question for the 3D 4-state Potts model,
we tried to fit the physical quantities to the standard expression, e.g.,
$a_\chi L^{\gamma/\nu}[1+b_\chi L^{-\omega} + \dots ]$, including a
sub-dominant correction-to-scaling term.
Since 4-parameter non-linear fits are notoriously unstable, we performed linear
fits where the exponents are kept fixed and only the amplitudes are
free parameters. In Fig.~\ref{FigChi2}, we show a 3D plot of the total
$\chi^2$ for the susceptibility fits as a function of $\gamma/\nu$ and $\omega$.
We observe a clear, stretched valley which confirms that $\gamma/\nu$ is close 
to $1.5$, but obviously this does not allow any reliable estimation of the
correction-to-scaling exponent $\omega$. The same procedure for $1/\nu$ gives 
qualitatively the same picture and confirms our previous estimate of
$1/\nu = 1.33$. For $\beta/\nu$, on the other hand, the $\chi^2$-landscape
turns out to be very flat and extremely sensitive to the fit range.

To conclude, from large-scale MC simulations of the 3D 4-state bond diluted  
Potts model we obtained a) the phase diagram in the $p-T$ plane in very good 
agreement with (rescaled) mean-field theory, b) the approximate
location of the tricritical point around $p_{TCP}=0.76(8)$, and 
c) at the dilution $p = 0.56$ 
clear evidence for softening of the rather strong first-order phase transition
in the pure case  towards a continuous transition
with estimates for the critical exponents of
$\nu=0.752(14)$, $\gamma/\nu=1.50(2)$, $\gamma=1.13(4)$, $\beta/\nu=0.65(5)$,
and $\beta=0.49(5)$. These
are clearly different from the values for both the disordered 3D Ising  
[$\nu=0.684(5)$, $\gamma/\nu=1.963(5)$] 
 and
the  $3-$state Potts
[$\nu=0.690(5)$, $\gamma/\nu=1.922(4)$] models.


\vspace*{-0.4cm}
\acknowledgments
We gratefully acknowledge financial support by the DAAD and EGIDE
through the PROCOPE exchange programme. C.C. thanks the DFG for financial
support 
through the Graduiertenkolleg ``Quantenfeldtheorie'' in
Leipzig. Work partially supported by the computer-time grants hlz061
of NIC, J\"ulich, C2000-06-20018 of the Centre Informatique National
de l'Enseignement Sup\'erieur (CINES), and C2000015 of the Centre de
Ressources Informatiques de Haute Normandie (CRIHAN).

\def\paper#1#2#3#4#5{#1, #3 {\bf #4}, \rm #5 (#2).}
\def\papers#1#2#3#4#5{#1, #3 {\bf #4}, \rm #5 (#2)}
        \def\PRB{Phys. Rev. B}
        \def\PRE{Phys. Rev. E}
        \def\PRL{Phys. Rev. Lett.}
        \def\JPA{J. Phys. A: Math. Gen.}

\end{multicols}

\end{document}